\documentclass[prl,onecolumn,showpacs,draft,preprint]{revtex4}%
\usepackage{amsmath}
\usepackage{graphicx}
\usepackage{amsfonts}
\usepackage{amssymb}%
\begin{document}
\title{Two-dimensional correlation spectroscopy of two-exciton resonances in
semiconductor quantum wells}
\date{\today{}}
\author{Lijun Yang and Shaul Mukamel}
\affiliation{Chemistry Department, University of California, Irvine, CA 92697-2025, United States}

\begin{abstract}
We propose a three-pulse coherent ultrafast optical technique that is
particularly sensitive to two-exciton correlations. \ Two Liouville-space
pathways for the density matrix contribute to this signal which reveals double
quantum coherences when displayed as a two-dimensional correlation plot.
Two-exciton couplings spread the cross peaks along both axes, creating a
characteristic highly resolved pattern. This level of detail is not available
from conventional one-dimensional \ four-wave mixing or other two-dimensional
correlation spectroscopy signals such as the photo echo, in which two-exciton
couplings show up along a single axis and are highly congested.

\end{abstract}

\maketitle

Investigating the correlations of multiple excitons in semiconductors is a
challenging many-body problem that had drawn considerable theoretical
\cite{H-K,AxtRMP,Rossi,Osterich256,Schafer161}and experimental
\cite{Shahbook,Chemla} attention. \ Correlations of two-excitons beyond
Hartree-Fock (HF) may result in either a red-shift (two-exciton binding
energy,TBE) or a blue-shift (two-exciton scattering energy, TSE). \ In GaAs
semiconductor quantum wells both couplings are a few meVs and may not be
easily resolved. \ Two-exciton formation has been reviewed in
Refs.\cite{Shahbook,Kalt,AxtRev}. \ Coherent ultrafast four-wave mixing (FWM)
\cite{Shahbook} provides a direct probe for two-exciton correlations in
quantum wells. \ The best evidence for bound two-excitons and the most
accurate TBE in GaAs quantum wells are obtained by time integrated FWM
(TIFWM), where signals are displayed as a function of a single (time or
frequency)\ variable.\cite{Shahbook,Wang1,Mayer237} Quantum beats were
observed in the negative-delay two-pulse signal along $2\mathbf{k}%
_{b}-\mathbf{k}_{a}$ where $\mathbf{k}_{b}$ arrives first, and in the
positive-delay three-pulse signal along $\mathbf{k}_{a}+\mathbf{k}%
_{b}-\mathbf{k}_{c}$ ($\mathbf{k}_{c}$ comes after $\mathbf{k}_{b}$). \ Many
attempts have been made to enhance the spectral resolution by displaying FWM
signals versus two time variables.\cite{KochM,Cundiff96PRL,Euteneuer} \ 

Multidimensional analysis of coherent signals is commonly used in NMR to study
correlations between spins.\cite{Ernst} \ These techniques were recently
extended to the femtosecond regime\cite{paper386} and applied to several
chemical and biological systems.\cite{Hochstrasser,Fleming} \ Three ultrashort
laser pulses generate a signal which is heterodyne-detected by a fourth pulse
in a chosen phase-matching direction (Fig. 1 top left). \ Three time delays
($t_{1},t_{2},t_{3}$) can be controlled between the chronologically ordered
pulses, $\mathbf{k}_{1}$, $\mathbf{k}_{2}$, $\mathbf{k}_{3}$ and the
heterodyne pulse $\mathbf{k}_{s}$. \ For an excitonic system the signal can be
generated along the phase-matching directions, $-\mathbf{k}_{1}+\mathbf{k}%
_{2}+\mathbf{k}_{3}$, $\mathbf{k}_{1}-\mathbf{k}_{2}+\mathbf{k}_{3}$ and
$\mathbf{k}_{1}+\mathbf{k}_{2}-\mathbf{k}_{3}$. \ We shall denote these three
techniques as $S_{\mathbf{I}}$, $S_{\mathbf{II}}$ and $S_{\mathbf{III}}$
\cite{298,paper487} respectively. \ The time-domain nonlinear response is
given by combinations of multi-time correlation functions which depend on the
time delays $t_{1}$, $t_{2}$ and $t_{3}$.\ \ Displaying the signal as a
function of two time delays (or their conjugate frequencies) while holding the
third fixed gives the two-dimensional correlation spectroscopy (2DCS) signals.
\ Note that $t_{1}$,\ $t_{2}$\ and $t_{3}$\ are positive. \ This is different
from conventional FWM where there is no fixed time ordering among the
$\mathbf{k}_{a}$, $\mathbf{k}_{b}$, $\mathbf{k}_{c}$ pulses and their delays
can be either positive or negative. \ We shall show that controlling the time
ordering in 2DCS is critical for the unambiguous identification of many-body
correlations. \ $S_{\mathbf{I}}$ (photon echo) signals were recently reported
in GaAs heterostructures\cite{Cundiff06PRL,Yang2DCS,Kuznetsova,Langbein}. \ In
this work, we demonstrate that the $S_{\mathbf{III}}$ technique, when
displayed in a proper projection plane, can access electronic correlations
beyond HF such as bound and unbound two-excitons with very high resolution,
not available in 1D FWM\ and any other 2DCS techniques. \ 

We employ a multi-band 1D tight-binding
Hamiltonian\cite{Sieh456PRL,Meierrecentbook} for calculating the
$S_{\mathbf{III}}$ signal. \ It includes a simplified Coulomb interactions and
accounts for the heavy-hole (HH), light-hole (LH)\ excitons and their
continuum states in a tractable way. \ This model has been successfully used
to reproduce many 1D FWM\cite{Meierrecentbook} and the recent $S_{\mathbf{I}}$
2DCS\cite{Kuznetsova} experiments. \ For clarity we focus on the HH\ spectral
region. \ With near-resonant excitations, only two terms contribute to the
third-order optical response\ in the phase-matching direction $\mathbf{k}%
_{1}+\mathbf{k}_{2}-\mathbf{k}_{3}$. \ These are represented by the
double-sided Feynman diagrams shown in the bottom of Fig. 1(left), where\ $e$
and $f$ represent respectively HH exciton and two-exciton manifolds and
$g$\ represents ground state (top right of Fig. 1)$\ $The rules of these
diagrams are given in Ref. \cite{298}. \ 

During $t_{1}$ and $t_{2}$, both pathways are respectively in a single
$\left(  \rho_{eg}\right)  $ and double $\left(  \rho_{fg}\right)  $ quantum
coherences. \ They differ only in the third interval $t_{3}$ ($\rho_{eg}$ for
(i) and $\rho_{fe}$ for (ii)). \ Signatures of two-exciton couplings thus show
up along $t_{3}$ and $t_{2}$ but not $t_{1}$. \ To enhance the resolution, we
therefore look at $\left(  t_{3},t_{2}\right)  $ correlation plots. \ We shall
display the signal by a double Fourier transform with respect to $t_{3}$ and
$t_{2}$ with the corresponding conjugate frequencies $\Omega_{3}$ and
$\Omega_{2}$, holding $t_{1}$ fixed.$\ $\ The expected 2DCS, $\mathbf{S}%
_{\mathbf{III}}\left(  \Omega_{3},\Omega_{2},t_{1}\right)  $, is schematically
sketched at the bottom of Fig. 1. \ The individual contributions of pathways
(i) and (ii) are denoted by the symbols given at the bottom of each Feynman
diagram. $f$ can be either red-shifted (bound) $f=2e-\Delta_{_{b}}$,
un-shifted $f=2e$ or blue-shifted (scattering) $f=2e+\Delta_{_{s}}$ relative
to twice the HH\ exciton energy, $2e$. \ Red(blue) energy shift corresponds to
two correlated excitons with opposite(same) spins. \ Open circles represent
un-shifted two-excitons and solid (open) symbols represent red(blue) shifted
two-excitons. \ Energy shifts from dressed states are negligible in the
present weak-field $\chi^{\left(  3\right)  }$ limit and thus the HF
(mean-field) contribution gives energy-unshifted two-excitons. \ 

We now examine the five peaks in Fig. 1. \ We note that the peaks resulting
from (i), are only spread along $\Omega_{2}$ and form a vertical pattern since
it has $eg$ resonances along $\Omega_{3}$, while for (ii) the peaks are spread
along both axes (diagonal pattern). \ Four properties make this signal
particularly useful for resolving various two-excitons and even the detailed
structure of the two-exciton continuum. \ First, along $t_{2}$ there are only
$fg$ resonances and thus the $\Omega_{2}$\ axis provides a clean projection
for two-excitons, without interference from $eg$ resonances. $\ $Second, the
two-exciton correlation energy can be obtained from a single peak.
\ $\Delta_{_{b}}$ is extracted from the solid square at $\left(
e,2e-\Delta_{_{b}}\right)  $ and $\Delta_{_{s}}$ from the open square at
$\left(  e,2e+\Delta_{_{s}}\right)  $. \ In TIFWM, in contrast, $\Delta_{_{b}%
}$ is determined by subtracting the energies of two peaks. \ However, the HF
peak (the circle) is not always clearly visible for certain pulse
polarizations and laser detunings. \ Third, the squares from (i) for different
$f$\ are generally much stronger than triangles from (ii). \ Thus it is
sufficient to get the two-exciton energies from the well-resolved and strong
(i) peaks (squares) along $\Omega_{2}$\ even when the triangles are not
resolved. \ Fourth, both axes $\Omega_{2}$ and $\Omega_{3}$ involve
two-exciton resonances. $\ $Along $\Omega_{2}$, there are only $fg$
resonances, while along $\Omega_{3}$ there is both $eg$ (i) and $fe$ (ii).
$\ $Spreading two-exciton resonances along both axes is critical for the high
resolution of two-excitons. \ Along $\Omega_{3}$ the two-exciton contribution
($fe$ in (ii) ) is weaker than the single-exciton one ($eg$ in (i)) and thus
may not be easily resolved. \ However, we shall show later that even this weak
signature along $\Omega_{3}$ is crucial for achieving high resolution when
combined with the information from the $\Omega_{2}$ projection. \ 

Using the tight-binding Hamiltonian\cite{Sieh456PRL,Meierrecentbook} with 10
sites, we have implemented the $\chi^{\left(  3\right)  }$\ formalism for
closing the infinite hierarchy of equations that has been widely applied to
Frenkel\cite{spano1,Leegwater} and Wannier\cite{AxtRMP,Axt94} excitons. \ In
Fig. 2, we present the simulated 2D spectra obtained by solving Eqs. 20 and 21
of Ref.\cite{Yang2DCS}) using periodic boundary conditions. \ Using the same
parameters of Ref. \cite{Weiser}, we set the dephasing times for excitons
$\tau_{ex}=2$\ ps and two-excitons $\tau_{2ex}=1$\ ps in panels A-E. \ Panel E
is obtained by cross-circular Gaussian pulses, where $\mathbf{k}_{s}$ is
linearly polarized and $\mathbf{k}_{3}$, $\mathbf{k}_{2}$ and $\mathbf{k}_{1}$
are respectively right, left and right circularly polarized. \ All other
panels are calculated with co-linear Gaussion pulses. \ Pulse parameters are
chosen such that red shifted two-excitons are selectively excited. \ The
optical power spectra width, $\Delta_{opt}$, and carrier frequency,
$\omega_{c}$(relative to $e$) are given in the caption. \ 

In all panels, the origin is $e$ for $\Omega_{3}$ and $2e$\ for $\Omega_{2}$.
\ All peaks or shoulders are assigned using the symbols given in Fig. 1.
\ Circle and solid square are strong and well resolved along $\Omega_{2}$,
while solid square and triangle are not resolved along $\Omega_{3}$. \ The
$\Omega_{2}$ value of solid square gives the TBE, $\Delta_{_{b}}$. \ Panels A
and B are calculated respectively with and without correlated two-excitons.
\ As expected, features related to red shifted two-excitons such as the solid
square and triangle, disappear. \ Panels C and D repeat the calculations of A
and B for pulses tuned to the blue-shifted two-excitons. \ One can obtain
blue-shifted TSE, $\Delta_{_{s}}$, from the open triangle (a shoulder but not
a resolved peak) in panel C, even though we can not resolve open square from
two-exciton continuum. \ In panel E (cross-circular excitation), the HF
contribution (the circle) at the origin ($e,2e$)=($0,0$) is missing.
\ However, one can still obtain the TBE using the solid square. \ We also note
that TBE may be extracted despite the large line broadening. \ Panel F shows
the 2D spectrum calculated with faster dephasing times $\tau_{ex}=1.0$ ps and
$\tau_{2ex}=0.5$ ps. \ Even though we can no longer resolve circle and solid
square along the clean $\Omega_{2}$ axis, it is still possible to obtain their
energies from the 'ridge' of the contour lines (two white lines). \ The TBE is
given by the $\Omega_{2}$ value of the bottom line. \ 

The TBE can be obtained from the intense and well-resolved solid square in
panel A. We thus circumvent the difficulty of probing TBE from the splitting
between solid square and the unresolved solid triangle, which overlap with
$eg$ and $fe$\ resonances along $t_{3}$. \ Additional calculations also show
that the TBE related solid square always has a solid triangle to the red, and
the TSE related open square always has a open triangle to the blue, as
predicted in Fig. 1. \ Moreover, the solid square and triangle (open square
and triangle) always appear in pairs and thus we can easily obtain accurate
TBE(TSE) by identifying both features, even when line broadening is large
(e.g. panel F). \ The much weaker solid triangle from pathway (ii) plays a
crucial role in identifying the solid square in panel F. \ Similarly, in panel
C we can not identify the open square without the help of the open triangle.
\ Thus it is impossible to resolve the signature of blue-shifted two-excitons
(unresolved open square and triangle) in panel C with conventional 1D FWM
where the signature will either be covered by the two-exciton continuum along
$\Omega_{2}$ or along $\Omega_{3}$ by the broadening of the much stronger
single-exciton peak, the circle. \ This signature can not be obtained from any
other 2DCS techniques where two-excitons show up along a single axis.
\ However, with the panoramic 2D view offered by $\mathbf{S}_{\mathbf{III}}$,
we can easily obtain very accurate TSE, even though the open square and
triangle are weak may not be resolved along any single axis. \ 

In $\mathbf{S}_{\mathbf{I}}$, we had demonstrated the first partial separation
of two-excitons\cite{Yang2DCS} along $\Omega_{3}$ due to the overlapping $eg$
and $fe$\ resonances along $t_{3}$, as shown by the elongation of peaks in
both experiments and simulations.\cite{Yang2DCS,Kuznetsova} \ $\mathbf{S}%
_{\mathbf{III}}$ provides an additional separation of two-excitons by
spreading them along $\Omega_{2}$. \ Without this separation, one can not
resolve any two-exciton features in panel F, let alone the accurate TBE. \ It
is the combination of the two dimensions $\Omega_{3}$ and $\Omega_{2}$ that
makes it possible to go from two ambiguous, unresolved open square and
triangle in panel C, to retrieve the unique signature of blue-shifted
two-excitons and obtain the TSE. \ 

The $\mathbf{S}_{\mathbf{III}}$ technique provides a new perspective into the
capacity of TIFWM experiments to provide TBE.\cite{Wang1,Mayer237} \ Let us
recast the existing TIFWM signal using 2DCS terminology. \ The two-pulse
signals\cite{Wang1} along $2\mathbf{k}_{b}-\mathbf{k}_{a}$ is given by
$W_{A}\left(  t_{1}\right)  =\int_{-\infty}^{+\infty}\left\vert S_{\mathbf{I}%
}\left(  t_{3},t_{2}=0,t_{1}\right)  \right\vert ^{2}dt_{3}$ (positive delay)
and $W_{B}\left(  t_{2}\right)  =\int_{-\infty}^{+\infty}\left\vert
S_{\mathbf{III}}\left(  t_{3},t_{2},t_{1}=0\right)  \right\vert ^{2}dt_{3}$
(negative delay). \ Three-pulse TIFWM signals\cite{Mayer237} along
$\mathbf{k}_{a}+\mathbf{k}_{b}-\mathbf{k}_{c}$ correspond to $W_{C}\left(
t_{2}\right)  =\int_{-\infty}^{+\infty}\left\vert S_{\mathbf{III}}\left(
t_{3},t_{2},t_{1}=t_{1}^{0}\right)  \right\vert ^{2}dt_{3}$ (positive delay)
and $W_{D}\left(  t_{2}\right)  =\int_{-\infty}^{+\infty}\left\vert
S_{\mathbf{II}}\left(  t_{3},t_{2},t_{1}=t_{1}^{0}\right)  \right\vert
^{2}dt_{3}$ (negative delay). \ Strong quantum beats only show up in $W_{B}$
and $W_{C}$, both related to $S_{\mathbf{III}}$ and depend on $t_{2}$. \ This
is clear from our pathway analysis. \ First, along $t_{2}$\ there are only
two-exciton resonances ($fg$) as shown in diagrams (i) and (ii) of Fig. 1, and
thus we have well-defined quantum beats along $t_{2}$ and well-resolved peaks
along $\Omega_{2}$. \ Second, the squares from (i) show up mainly as
$eg$\ resonances along $t_{3}$ and can be much stronger than the triangles,
although they are all induced by correlated two-excitons. Without such
correlations, there will only be a circle at the origin. \ It is the stronger
solid square and circle that give quantum beats. \ However, there are no
appreciable beats for $W_{A}$ and $W_{D}$. \ Fig. 3 shows the pathways for
$S_{\mathbf{I}}$ and the schematic 2D spectra projected respectively onto
$\left(  \Omega_{3},-\Omega_{1}\right)  $ and $\left(  \Omega_{3},\Omega
_{2}\right)  $. \ Obviously, there are no two-exciton splitting along either
$\Omega_{1}$ or $\Omega_{2}$ and thus $S_{\mathbf{I}}$ does not show quantum
beats with respect to $t_{1}$ or $t_{2}$, as in $W_{A}$. \ A similar
conclusion applies to $S_{\mathbf{II}}$, which corresponds to $W_{D}$. \ The
only axis along which quantum beats occur for $S_{\mathbf{I}}$ and
$S_{\mathbf{II}}$ is $t_{3}$. \ However, it is hard to observe such beats with
any technique (including $S_{\mathbf{III}}$), because there are always
dominant $eg$ resonances along $t_{3}$ for all three 2DCS techniques and thus
the two-exciton\ resonances are not well resolved. \ 

$\mathbf{S}_{\mathbf{III}}$ also demonstrates an important limitation of TIFWM
in probing TBE. \ Our calculations show that TIFWM only yields the correct TBE
in ideal cases such as Fig. 2A where the $\Omega_{2}$ coordinate of the circle
coincides with $2e$ and its intensity is comparable to solid square.
\ However, the $\Omega_{2}$ value and intensity of the circle are very
sensitive to the pulse width, detunings, pulse polarizations, and are further
affected by the cancellation of two-exciton continuum to the HF portion.
\ Calculations with long dephasing times (not shown) also suggest that the
two-exciton continuum is far from uniform. \ Although the solid square has a
fixed $\Omega_{2}$, its intensity is also very sensitive to the pulse
properties. \ Therefore the circle and solid square do not always dominate the
signal. \ Other components from two-exciton continuum or even open square may
also contribute or even dominate the quantum beats. \ This makes the beating
experiment less accurate for probing TBE. \ Furthermore, in the most favorable
case for probing TBE where only bound two-excitons are generated (e.g. the
cross-circular excitations in Fig. 2E), the HF peak (circle) disappears
completely. \ In this case, any beating frequency obtained is not connected to
TBE. \ Therefore, one has to use other pulse polarizations to obtain quantum
beats where the position and intensity of the circle is affected by other
types of two-excitons. \ 

The advantage of $\mathbf{S}_{\mathbf{III}}$ is even more pronounced when HH
and LH excitons both contribute and TIFWM signal is congested with many types
of beating components along $t_{2}$ arising from pure HH (LH) and mixed
two-excitons. \ By dispersing all pathways into an extra dimension and
spreading them along $\Omega_{2}$ over a spectral range twice that of
$\mathbf{S}_{\mathbf{I}}$\ and $\mathbf{S}_{\mathbf{II}}$ techniques,
$\mathbf{S}_{\mathbf{III}}$ achieves a complete separation of pure HH (LH)
two-excitons from the mixed two-excitons. \ Furthermore, by choosing specific
pulse polarizations, the contribution from correlated pure or mixed
two-excitons can be completely isolated from their corresponding HF
contributions. \ 

In summary, we propose a 2DCS technique, $S_{\mathbf{III}}$, that is
intrinsically sensitive to two-exciton correlations and can distinguish
between different species of excitons and two-excitons in photo-excited
semiconductors by spreading them along two axes. \ Combining the information
from different Liouville space pathways,\cite{298} $S_{\mathbf{III}}$ can
accurately retrieve TBE even when it is smaller than exciton line broadening.
\ We further report an unambiguous signature of correlated unbound
two-excitons lying underneath the two-exciton continuum. \ Conventional 1D or
2D techniques can not achieve such resolution for the lack of time ordering
among pulses, which is essential for the identification of specific pathways.
\ For example, either 1D or 2D pump-probe signals generally mix the
$S_{\mathbf{I}}$ and $S_{\mathbf{II}}$ techniques and thus one can not employ
them to identify different pathways, let alone combine them to give higher
resolution. \ $S_{\mathbf{III}}$ also provides new insights into the quantum
beats in the most accurate TIFWM for obtaining TBE\cite{Wang1,Mayer237} of
quantum wells, which are dominated by one of the two pathways and generally do
not provide accurate TBE. 

$S_{\mathbf{III}}$ could be applied to other strongly correlated systems such
as Bose-Einstein condensate and superconductors. \ For instance, it could be
helpful in the considerable current effort aimed at the coherent manipulation
of excitons and exciton Bose-Einstein condensation. \ Two-exciton effects are
critical for achieving these goals. \ 

\begin{acknowledgments}
The support of the Chemical Sciences, Geosciences, and Biosciences Division,
Office of Basic Energy Sciences, U. S. Department of Energy is gratefully acknowledged.
\end{acknowledgments}

Figure captions

Fig. 1 \ Top: Schematic experimental setup (left) and exciton level scheme
(right). Bottom: Feynman diagrams of $S_{\mathbf{III}}$ technique (left) and
schematic 2D spectrum (right). 

Fig. 2 \ (Color online) Calculated $\mathbf{S}_{\mathbf{III}}(\Omega
_{3},\Omega_{2},t_{1})$ for (A) and (B), $\Delta_{opt}=3.9$ meV, $\omega
_{c}=-2.85$ meV; (C) and (D), $\Delta_{opt}=1.8$ meV, $\omega_{c}=2.15$ meV;
\ (E), $\Delta_{opt}=3.9$ meV, $\omega_{c}=-3.85$ meV; \ and (F) $\Delta
_{opt}=3.9$ meV, $\omega_{c}=-1.85$ meV. \ 

Fig. 3 \ (Color online) Top: Feynman diagrams of $S_{\mathbf{I}}$ technique.
\ Bottom: Schematic 2D spectra, $S_{\mathbf{I}}\left(  \Omega_{3}%
,t_{2},-\Omega_{1}\right)  $ (left), $S_{\mathbf{I}}\left(  \Omega_{3}%
,\Omega_{2},t_{1}\right)  $ (right). \

\end{document}